\documentclass[manuscript]{aastex63}

\received{}
\revised{}
\accepted{}
\submitjournal{ApJ}

\shorttitle{Causal interaction between the flow and magnetic fields}
\shortauthors{Inceoglu et al.}

\begin{document}

\title{Causal Interaction between the subsurface rotation rate residuals and radial magnetic field in different timescales}

\correspondingauthor{Fadil Inceoglu}
\email{fadil@gfz-potsdam.de}

\author[0000-0003-4726-3994]{Fadil Inceoglu}
\affiliation{GFZ German Research Centre for Geosciences, Telegrafenberg, 14473, Potsdam, Germany}

\author[0000-0002-3834-8585]{Rachel Howe}
\affiliation{School of Physics and Astronomy, University of Birmingham, Edgbaston, Birmingham B15 2TT, UK}
\affiliation{Stellar Astrophysics Centre (SAC), Department of Physics and Astronomy, Aarhus University, Ny Munkegade 120, DK-8000 Aarhus C, Denmark}

\author{Paul T. M. Loto'aniu}
\affiliation{Cooperative Institute for Research in Environmental Sciences, University of Colorado Boulder, Boulder, CO, USA}
\affiliation{National Centers for Environmental Information, National Oceanographic and Atmospheric Administration, Boulder, CO, USA}




\begin{abstract}

We studied the presence and spatiotemporal characteristics and evolution of the variations in the differential rotation rates and radial magnetic fields in the Schwabe and Quasi-biennial-oscillation (QBO) timescales. To achieve these objectives, we used rotation rate residuals and radial magnetic field data from the Michelson Doppler Imager on the Solar and Heliospheric Observatory and the Helioseismic and Magnetic Imager on the Solar Dynamics Observatory, extending from May 1996 to August 2020, covering solar cycles 23 and 24, respectively. Under the assumption that the radial surface magnetic field is non-local and the differential rotation is symmetric around the equator, our results suggest that the source region of the Schwabe cycle is confined between $\sim$30$^{\circ}$ N and S throughout the convection zone. As for the source region of the QBO, our results suggest that it is below 0.78R$_{\odot}$.

\end{abstract}

\keywords{QBO, rotation rate, differential rotation, radial magnetic field, convergent cross mapping}


\section{Introduction} \label{sec:intro}

The Sun is a magnetically active variable star. The Sun governs the space-climate and space-weather throughout the heliosphere via its magnetic activity variations in short-, mid-, and long-term timescales. The most well-known magnetic activity cycle is the Schwabe cycle \citep{1844AN.....21..233S}, the period of which ranges from 9 to 13 years. The Schwabe cycle is superimposed on longer-term variations, such as $\sim$90-year Gleissberg \citep{1939Obs....62..158G} and $\sim$210-year Suess \citep{Suess1980} cycles. The Sun also shows shorter quasi-periodic variations that are 160-day Rieger-type periodicities \citep{1984Natur.312..623R} and quasi-biennial oscillations (QBOs), the period of which ranges from 0.6 to 4 years. It was also pointed out that there is a clear separation at 1.5 yr, indicating two groups of variations below and above this value \citep{2014SSRv..186..359B}.

The QBOs are shown to be more intermittent signals and the variations in their amplitude are in-phase with the Schwabe cycle, meaning they attain their highest (lowest) amplitude during the solar cycle maxima (minima) \citep{2014SSRv..186..359B}. Together with exhibiting signals over all solar latitudes \citep{2012ApJ...749...27V}, they are also shown to behave differently in each solar hemisphere \citep{2017ApJ...845..137G, 2019A&A...625A.117I}. The QBOs are found to be present from the subsurface layers to the surface of the Sun, and they can even be identified in the neutron counting rates measured on Earth as indicators of the Galactic Cosmic Ray intensities \citep{1998ApJ...509L..49B,2010SoPh..266..173K,2012A&A...539A.135S,2012ApJ...749...27V}. Recently, \citet{2021ApJ...920...49I} showed that the rotation rate residuals also show QBO and their amplitude increase with increasing depth. Therefore, the QBOs are thought to be global phenomena extending from the subsurface layers of the Sun to the Earth.

Among the physical mechanisms that were proposed to explain the existence and spatio-temporal behaviors of the QBOs, we can include spatio-temporal fragmentation of radial profiles of the rotation rates \citep{2013ApJ...765..100S}, 180$^{\circ}$ shifting of the active longitudes \citep{2003A&A...405.1121B}, a secondary dynamo operating in the subsurface shear layer at 0.95R$_{\odot}$ \citep{1998ApJ...509L..49B}, instability of the magnetic Rossby waves in the tachocline \citep{2010ApJ...724L..95Z}, and tachocline nonlinear oscillations (TNOs) through periodic energy exchange between the Rossby waves, differential rotation, and the present toroidal field \citep{2018ApJ...853..144D}. In addition, based on results from a fully nonlinear flux transport dynamos, \citet{2019A&A...625A.117I} proposed that there are indications for the QBOs to be generated via interplay between the flow and magnetic fields, where the turbulent ${\alpha}$-mechanism working in the lower half of the solar convection zone, which extends from 0.70R$_{\odot}$ to the surface. The bottom of the convection zone overlaps with the region of strong radial shear. Above this region a differential rotation pattern that depends strongly on latitude takes place \citep{2009LRSP....6....1H}.

Are the interactions between the magnetic and flow fields, as investigated by rotation rate residuals and radial magnetic fields, in the Schwabe and QBO timescales different at different depths and latitudes? Are there any preferred locations for the generation of these cyclic variations in the Sun's magnetic activity levels? To answer these questions, we utilized lagged cross correlation and convergent cross mapping analyses using surface magnetic field and subsurface flow field data from the Michelson Doppler Imager (MDI) on the Solar and Heliospheric Observatory (SOHO) and the Helioseismic and Magnetic Imager (HMI) on the Solar Dynamics Observatory (SDO), covering Solar Cycles 23 and 24. 

\section{Data} \label{sec:data}

We calculated the rotation rates based on regularized least squares (RLS) code using frequencies
derived from MDI/SOHO and HMI/SDO data. The rotation rates between May 1996 and February 2011 are calculated based on 59 sets of rotational splittings from 72-day spectra of MDI observations, while for the period between May 2010 to August 2020 53 sets from HMI. To splice the two data sets, we calculated an offset using the mean difference between the HMI and MDI rotation profiles over the 5 periods, where they overlap. This offset was applied because data from the MDI might be influenced by some systematic effects, which do not influence data from the HMI \citep{2013ApJ...767L..20H}.

The surface magnetic field data for the period spanning from Carrington Rotation (CR) 1909 (May 1996) to 2104 (December 2010) are calculated using MDI radial magnetic field synoptic charts \citep{1995SoPh..162..129S}, while for the period from CR 2097 (June 2010) to CR 2235 (October 2020) we used HMI's radial magnetic field synoptic charts \citep{2012SoPh..275..207S}. First of all we converted signed magnetic field strengths into unsigned magnetic field strength by simply taking the absolute values in each synoptic map per CR and then averaged them in time per latitude. To merge the two data sets, we re-scaled the HMI data using relationships given in \citet{2012SoPh..279..295L}.

Following to that, to have same temporal and spatial resolution, we have interpolated the merged rotation rate residual data using cubic spline method in 2D.

\section{Analyses and Results} \label{sec:analyses_res}

To study whether there are differences in interactions between the magnetic and flow fields in the Schwabe and the QBO timescales at different depths of the Sun, we merged the MDI/SOHO and HMI/SDO synoptic maps for radial magnetic fields spanning the last two solar cycles. In addition, we calculated the rotation rate residuals based on regularized least squares (RLS) code using frequencies derived from MDI and HMI data. Furthermore, to remove the potential effects from the annual periodic variations caused by the Earth's orbital inclination and tilt of the solar rotation axis, we low pass filtered the data with a cut-off frequency of 1.5 year$^{-1}$ using a Butterworth filter of degree 5. The Butterworth filter is a maximally flat filter in the passband that avoids any distortion of the low frequency components of the signal. They can also be used as low-pass, high-pass, and band-pass filters \citep{1978JGR....83.5510R}.

\begin{figure*}
\begin{center}
{\includegraphics[width=5.5in]{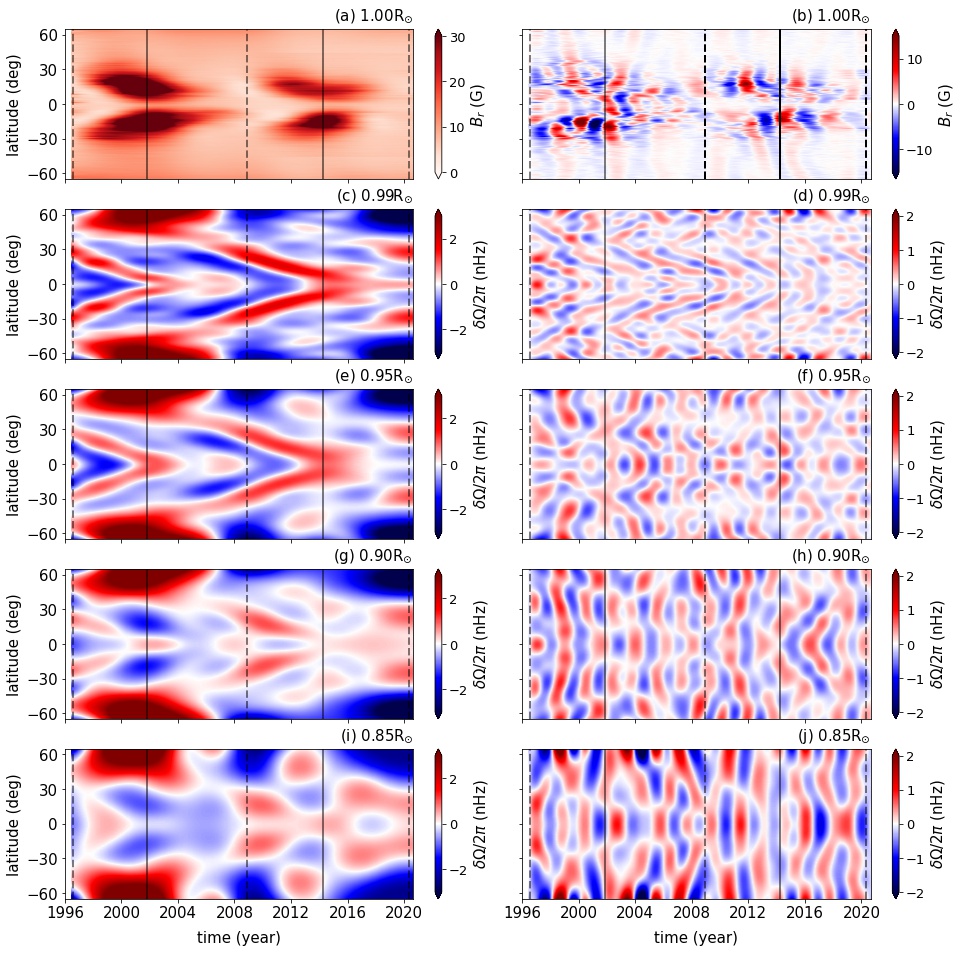}}
\caption{The left panels (a, c, e, g, and i) show low pass filtered average absolute magnetic field strength and rotation rate residuals covering solar cycles 23 and 24. The right panels (b, d, f, h, and j) display the band pass filtered average absolute magnetic field and rotation rate residuals covering solar cycles 23 and 24. Note that the limit of the color bar in Figure1b shows fluctuations around the mean and not signed magnetic field strength. Vertical dashed lines show solar cycle minima, while the vertical solid lines show solar cycle maxima. Also note that the rotation rate residuals are flipped around the equator.}
\label{fig:data}
\end{center}
\end{figure*}

The Schwabe cycle and QBOs have periods spanning between 9 – 13 years and 1.5 – 4 years, respectively. Therefore, to separate the average unsigned magnetic field strengths and rotation rate residuals into two different time scales, we used, once more, a Butterworth filter of degree 5 with a cutoff frequency of 4.5 year$^{-1}$. The Butterworth filter has been applied to each depth under consideration for the rotation rate residuals extending from $\sim$0.70R$_{\odot}$ to 1.00R$_{\odot}$ and to the surface average unsigned magnetic field strengths. We limited the latitudinal interval between 65$^{\circ}$ N and S because of data gaps in the surface magnetic field data due to the tilt of the rotation axis of the Sun as well as progressively decreasing reliability of the inversion with increasing latitude \citep{2009LRSP....6....1H}. We must note that the global helioseismic inversions cannot resolve the solar hemispheres, therefore the values are flipped around the equator for convenience. 

In the low pass filtered data, variations in the average unsigned magnetic field strengths in the Schwabe timescale can clearly be observed, where the magnetic field is stronger in the solar cycle 23 than that in solar cycle 24 (Figure~\ref{fig:data}a). We also show variations in the rotation rate residuals in the Schwabe timescale for selected solar depths of 0.99R$_{\odot}$, 0.95R$_{\odot}$, 0.90R$_{\odot}$, and 0.85R$_{\odot}$ in Figures~\ref{fig:data}c, e, g, and i, respectively. The rotation rate residuals in this timescale show faster-than-average and slower-than-average flow bands in each depth. There are pronounced differences in flow patterns in high latitudes above $\sim$45$^{\circ}$ between solar cycle 23 and 24. For example, at the solar cycle 23 maximum, there is a strong faster-than-average flow band above $\sim$45$^{\circ}$ latitude, whereas there is nothing similar at the solar cycle 24 maximum. During the declining phase of solar cycle 24 a slower-than-average flow band forms in the high latitudes, which cannot be observed during the same phase of solar cycle 23 (Figures~\ref{fig:data}c, e, g, and i). The lower latitudes below $\sim$45$^{\circ}$ latitude, on the other hand, show very similar behavior, having slower-than-average and faster-than-average flow bands form close to $\sim$45$^{\circ}$ latitude and propagate equator-ward throughout the solar cycles.

In the QBO timescale, the average unsigned magnetic field shows fluctuations around the mean value, which is generally confined between $\sim$35$^{\circ}$ N and S around the solar equator (Figure~\ref{fig:data}b). The rotation rate residuals in this time scale also show similar slower-than-average and faster-than-average flow bands distributed all latitudes throughout the solar cycles at each depth, except for those in 0.99R$_{\odot}$, which show more confined flow bands propagating equator-ward (Figures~\ref{fig:data}d, f, h, and j).

\subsection{Amplitude variations in the Schwabe and QBO timescales}

\begin{figure*}
\begin{center}
{\includegraphics[width=6in]{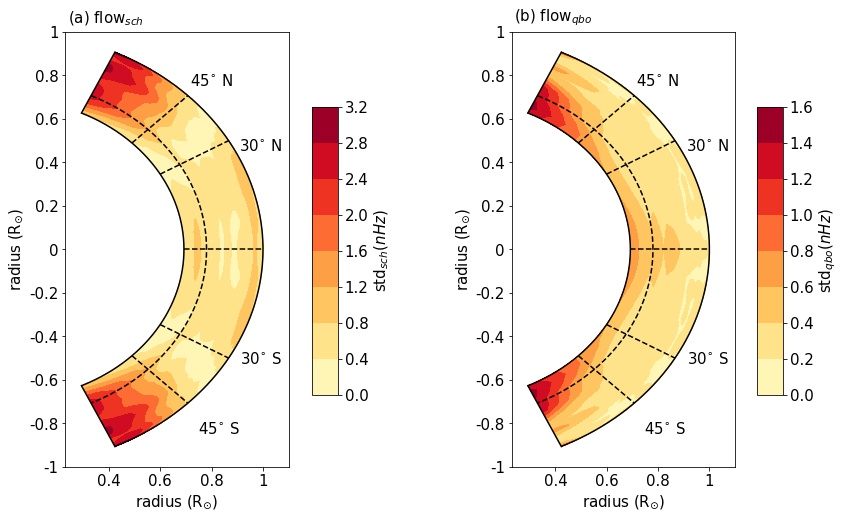}}
\caption{Amplitude of variations in the rotation rate residuals as a function of depth and latitude for the Schwabe timescale (a) and the QBO timescale (b). The dashed lines show 0.78R$_{\odot}$.} 
\label{fig:amplitude_flow}
\end{center}
\end{figure*}

We then calculate the amplitudes of the variations in the Schwabe and QBO time scales by simply calculating the standard deviations of variations at each latitude and depth ranging from 0.70$_{\odot}$ and 1.00$_{\odot}$ (Figures~\ref{fig:amplitude_flow}a, and b). The amplitude of variations in the Schwabe timescale increase with increasing latitude after $\sim$35$^{\circ}$ and it reaches its maximum around $\sim$60$^{\circ}$ latitude. The amplitude of variations in these higher latitudes show similar values down to around 0.78$_{\odot}$, after which they become weaker. For the region between 0$^{\circ}$ - $\sim$35$^{\circ}$ latitudes, the amplitude of the variations shows a decreasing trend down to around 0.80R$_{\odot}$ and it slightly increases (Figure~\ref{fig:amplitude_flow}a).

On the contrary, the amplitude of variations in the QBO timescales almost shows a reversed pattern. Although the latitudinal dependence of the amplitudes above $\sim$35$^{\circ}$ latitude is similar to those observed in the Schwabe timescale, the amplitude of variations reach their maximum below 0.78R$_{\odot}$. Below $\sim$35$^{\circ}$ latitude, on the other hand, the amplitude of variations increase with increasing depth, which is more pronounced around the solar equator (Figure~\ref{fig:amplitude_flow}b).

\begin{figure}
\begin{center}
{\includegraphics[width=2in]{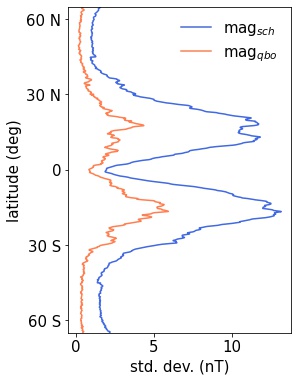}}
\caption{Amplitude of variations in the radial magnetic field as a function of latitude for the Schwabe timescale (blue) and the QBO timescale (orange).} 
\label{fig:amplitude_mag}
\end{center}
\end{figure}

Similar to the ration rate residuals, we also calculated the amplitudes of variations in the unsigned magnetic field strengths at each latitude for the Schwabe and QBO time scales (Figure~\ref{fig:amplitude_mag}). The two time scales show almost identical amplitude variations as a function of latitude; the amplitudes are higher in the magnetic activity bands that are confined between latitudes $\sim$40$^{\circ}$ N and S with maximum amplitudes are observed around $\sim$15$^{\circ}$ N and S latitudes. Interestingly, the amplitudes of variations in the Schwabe and QBO time scales decrease when we approach the solar equator from $\sim$15$^{\circ}$ N and S latitudes (Figure~\ref{fig:amplitude_mag}).

To investigate the interaction between the flow and magnetic fields, we had to make two assumptions for further analyses; (i) the rotation profiles are symmetric around the equator and (ii) the magnetic field measured on the surface is non-local in the convection zone, down to 0.71R$_{\odot}$, as information of the magnetic field strengths in below the surface of the Sun is not yet available. Earlier, it was shown that there are some asymmetries in the flow fields during the declining phase of cycle 23 and rising phase of solar cycle 24 at $\sim$0.99R$_{\odot}$ \citep{2018ApJ...861..121L}. Therefore, the results drawn from the following analyses must be approached with caution. The main idea here is to draw an average picture of the interactions between magnetic and flow fields.

\subsection{Lagged-Cross Correlations in the Schwabe and QBO timescales}

To investigate the linear relationship between the rotation rate residuals and average unsigned magnetic field, we use lagged-cross correlations at each depth and latitude.

\begin{figure*}
\begin{center}
{\includegraphics[width=6in]{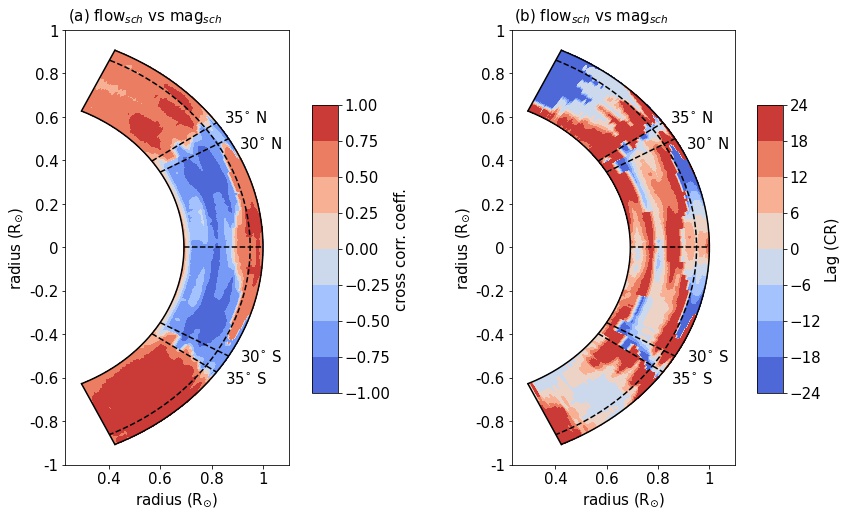}}
\caption{The highest lagged cross correlation coefficients (a) and the corresponding lags (b) between the rotation rate residuals (flow) and radial magnetic field (mag) in the Schwabe timescale. Note that the lag is denoted in Carrington Rotations (CR) and positive values indicate magnetic field leading the flow field, while negative lag means the flow field is leading the magnetic field. The dashed lines show 0.95$R_{\odot}$ to aid the eye (see text).} 
\label{fig:Xcorr_schw}
\end{center}
\end{figure*}

In the Schwabe timescale, there is a strong positive correlation in almost all depths at latitudes above $\sim$35$^{\circ}$ (Figure~\ref{fig:Xcorr_schw}a). In the both solar hemispheres the magnetic field is leading the flow field by 24 CRs at all depths in the latitudes between $\sim$35$^{\circ}$ and $\sim$45$^{\circ}$. After around $\sim$45$^{\circ}$ up to $\sim$60$^{\circ}$ latitude, although the there is still a positive correlation between the magnetic and flow fields, the flow field starts to lead the magnetic field with around 6 CRs. Above $\sim$60$^{\circ}$ latitude, the magnetic field leads the flow field the southern solar hemisphere, while on the contrary the flow field leads the magnetic field in the northern hemisphere (Figure~\ref{fig:Xcorr_schw}). Between $\sim$30$^{\circ}$ N and S around the solar equator and down to 0.95R$_{\odot}$, there is a high positive correlation between the magnetic and flow fields, where flow field leads the magnetic field with longer time with increasing latitude. However, between $\sim$5$^{\circ}$ N and S around the solar equator, magnetic field is still leading the flow field. Between $\sim$30$^{\circ}$ and $\sim$35$^{\circ}$ latitudes in the north and south, there is a strong inverse relationship between the radial magnetic field and rotation rate residuals at all depths and flow field is leading the magnetic field (Figure~\ref{fig:Xcorr_schw}). At depths below 0.95R$_{\odot}$ down to 0.71R$_{\odot}$, there is an inverse relationship between the flow and magnetic fields. This region also exhibits that most of the time, the magnetic field is leading the flow field (Figure~\ref{fig:Xcorr_schw}).

\begin{figure*}
\begin{center}
{\includegraphics[width=6in]{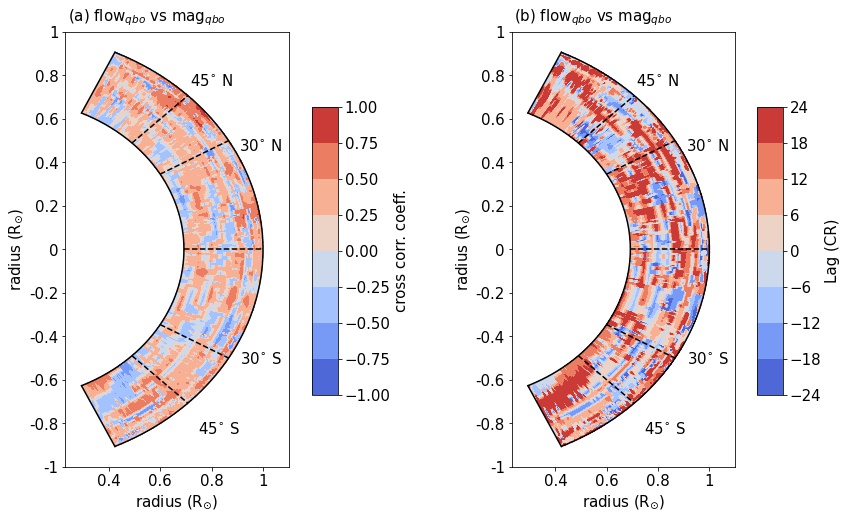}}
\caption{The same as Figure~\ref{fig:Xcorr_schw} but for the QBO time scale.} 
\label{fig:Xcorr_qbo}
\end{center}
\end{figure*}

In the QBO timescale, on the contrary, there is not a clear pattern in the relationship between the rotation rate residuals and radial magnetic field (Figure~\ref{fig:Xcorr_qbo}). One of the most pronounced feature that are not present in the Schwabe timescale is the patches of marginally strong positive correlations through the convection zone, down to 0.71R$_{\odot}$. These patches are observed to form in layers of positive and negative correlations. The magnetic field tends to lead the flow fields where the correlation coefficients are positive, while the flow field leads the magnetic field where the correlation is negative (Figure~\ref{fig:Xcorr_qbo}). 

\subsection{Causal relationship in the Schwabe and QBO timescales}

To study the non-linear causal relationship between the rotation rate residuals and average unsigned radial magnetic field, we used convergent cross mapping (CCM) method, which is first introduced by \citet{2012Sci...338..496S} in 2012. CCM is a novel method which can detect if two time-series originate from the same dynamical system based on measuring the predictability of one variable using the other \citep{2012Sci...338..496S}. The same dynamical system acts as a common attractor manifold leading to the two time-series to be causally linked. This case allows us to estimate the states of a causal variable using the affected variable. The overall predictive skill improves and converges with increasing time-series length. The key property which enables us to distinguish causation from simple correlation is this convergence criterion \citep{2012Sci...338..496S}. In this study, we use the {\it pyEDM}\footnote{URL: https://github.com/SugiharaLab/pyEDM} python library to study cause and effect relationship between the rotation rate residuals and radial unsigned magnetic field strengths as well as bidirectional influences between the two variables in each latitude and radius \citep{2012Sci...338..496S, 2016Sci...353..922Y}. To investigate the causal influence between the two variables, we first standardized each data set using their individual mean and standard deviation values, and then calculated their CCMs.

\begin{figure*}
\begin{center}
{\includegraphics[width=5in]{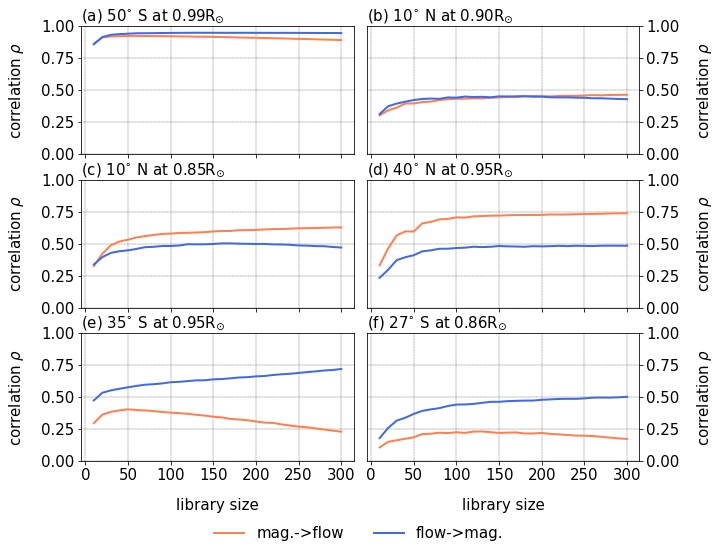}}
\caption{The direction of the causal interaction between the rotation rate residuals and radial magnetic field for the Schwabe (the left panel) and the QBO timescales (the right panel) for selected radius and latitudes. We give examples of where the directional causal influence cannot be determined (top panel), where there is a the stronger magnetic field influence on the flow field (middle panel), and stronger flow field influence on the mag field (bottom panel) cases.} 
\label{fig:ccm_sch_qbo_example}
\end{center}
\end{figure*}

To give an example of the analysis, we first show three different results for each timescale (Figure~\ref{fig:ccm_sch_qbo_example}). These results are for (i) where the causal influence of the magnetic field on the flow field is stronger  (orange), (ii) where the causal influence of the flow field on the magnetic field is stronger (blue), and (iii) where the direction of the causal influence cannot be determined (white) (Figure~\ref{fig:ccm_sch_qbo_example}). CCM results for the Schwabe and the QBO timescales show that the causal relationship between the flow and the magnetic field is bidirectional at every depth and latitude, which is expected considering back-reaction of the Lorentz Force on the flow field. However, the degree of these causal influences at each depth and latitude varies. For example, at 50$^{\circ}$ S and at 0.99R$_{\odot}$ in the Schwabe timescale and at 10$^{\circ}$ N and at 0.90R$_{\odot}$ in the QBO-time scale, the causal influence between the flow and magnetic fields is bidirectional with similar degrees (Figures~\ref{fig:ccm_sch_qbo_example}a and b), whereas the causal influence of the magnetic field on the flow field is stronger at 10$^{\circ}$ N and at 0.85R$_{\odot}$ and at 40$^{\circ}$ N and at 0.95R$_{\odot}$ in the Schwabe and the QBO-time scales, respectively (Figures~\ref{fig:ccm_sch_qbo_example}c and d). On the other hand, the causal influence of the flow field on the magnetic field is stronger at 35$^{\circ}$ S latitude and at 0.95R$_{\odot}$ in the Schwabe timescale and at 27$^{\circ}$ S and at 0.86R$_{\odot}$ in the QBO-time scale (Figures~\ref{fig:ccm_sch_qbo_example}e and f).

\begin{figure*}
\begin{center}
{\includegraphics[width=4.5in]{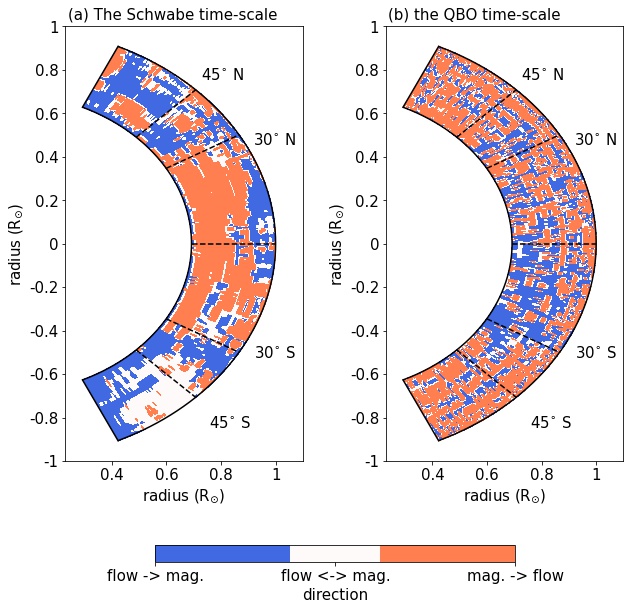}}
\caption{The direction of the causal interaction between the rotation rate residuals and radial magnetic field for the Schwabe (a) and the QBO timescales (b).} 
\label{fig:ccm_sch_qbo}
\end{center}
\end{figure*}

We then calculate the CCMs for each pair of flow and magnetic field data at each depth ranging from 0.71R$_{\odot}$ and 1.00R$_{\odot}$, and from 65$^{\circ}$ N to 65$^{\circ}$ S. We plotted the direction of causal influences found for the Schwabe and the QBO timescales to study whether there is a different pattern in causal relationship in different timescales (Figure~\ref{fig:ccm_sch_qbo}).

In the Shcwabe timescale, the causal influence of the magnetic field on the flow field is stronger and is confined between the latitudes of $\sim$30$^{\circ}$ N and S generally at all depths, except for the shallow region reaching down to 0.85$_{\odot}$ and between $\sim$15$^{\circ}$ N and $\sim$10$^{\circ}$ S latitudes. In this region, the causal influence of the flow field on the magnetic field is stronger (Figure~\ref{fig:Xcorr_schw}a). Another important feature is above $\sim$45$^{\circ}$ N and S latitudes, the direction of the causal influence switches generally to be from the flow field to the magnetic field with regions of bidirectional and reversed relationships observed in the S and N hemispheres, respectively. An interesting feature is seen between $\sim$45$^{\circ}$ and $\sim$60$^{\circ}$ N latitudes and between $\sim$0.75R$_{\odot}$ and $\sim$0.85R$_{\odot}$, where the causal effects of the magnetic field on the flow field is stronger. A similar pattern also exists for the southern hemisphere, however the degree of influence between the magnetic and flow fields  in this region is very similar (Figure~\ref{fig:Xcorr_schw}a).

The direction of the stronger causal effect in the QBO timescale, on the other hand, is mostly from the magnetic field to the flow field, at all latitudes and depths, with some small regions especially below  $\sim$0.85$_{\odot}$ where it shows opposite behavior (Figure~\ref{fig:ccm_sch_qbo}b). More specifically, the latitudes between 0 and $\sim$30$^{\circ}$ S, the direction of stronger causal effect is more from the flow field to the magnetic field, while the same latitudinal band in the northern solar hemisphere show that the effect of the magnetic field on the flow field is stronger (Figure~\ref{fig:ccm_sch_qbo}b). Above $\sim$30$^{\circ}$ N and S latitudes, the influence of the magnetic field on the flow field is stronger everywhere (Figure~\ref{fig:ccm_sch_qbo}b).


\section{Discussion and Conclusions} \label{sec:dis_conc}

The results point out that the rotation rate residuals and average unsigned magnetic field strengths show different patterns in two different timescales. The strong unsigned magnetic field is confined in a band between $\sim$5$^{\circ}$ and $\sim$30$^{\circ}$ in the northern and southern hemispheres, where they show  equator-ward propagation in the Schwabe timescale. In the QBO timescale, on the other hand, the magnetic field strengths fluctuates around the mean value creating lower than average and higher than average magnetic field bands, which show both equator-ward and poleward propagation. These results are in line with \citet{2012ApJ...749...27V} who showed that there are pole-ward and equator-ward propagation bands in the radial and meridional components of the magnetic fields in the QBO timescales during solar cycles 21 and 22. The explanation for the differences in behavior observed in the unsigned magnetic fields comes from the generation of the toroidal magnetic fields by the solar dynamo. In the Schwabe timescales, this behavior is closely related with the emergence of the active regions, while in the QBO timescale it resembles the pole- and equator-ward transportation of the residual magnetic field via flows after the bipolar active region cancels itself out \citep{2012ApJ...749...27V}. The amplitude of variations in the Schwabe and the QBO timescales are very similar, having higher amplitudes confined between $\sim$5$^{\circ}$ and $\sim$30$^{\circ}$ in the northern and southern hemispheres with the maximum amplitudes around $\sim$20$^{\circ}$ N and S.

The rotation rate residuals in the Schwabe timescale exhibited faster-than-average and slower-than-average flow bands that form $\sim$45$^{\circ}$ and propagate equator- and pole-ward. The slower-than-average flow bands generally coincide with regions where the magnetic field is stronger. This is a result of the back-reaction of the Lorentz Force on the flow field. We also observe a tail-like structure in the slower-than-average flow band, extending into the next cycle by around 2 years at the depths above $\sim$0.95R$_{\odot}$. The tail-like structure, indicating the overlapping period between two consecutive cycles, in the flow field can be explained by flux transport dynamos with Babcock-Leighton (BL) mechanism together with turbulent $\alpha$-effect operating throughout the convection zone are the main source for the generation of the poloidal field from a pre-existing toroidal field \citep{2014A&A...563A..18P, 2016ApJ...828...41S}. Solar dynamos that use only the BL mechanism as well as only thin-shell dynamos with turbulent $\alpha$-effect, on the other hand, tend to generate longer overlapping periods between the cycles \citep{2001ApJ...559..428D, 2005A&A...437..699D, 2006MNRAS.371..772B, 2017ApJ...847...69K, 2017ApJ...848...93I,2019A&A...625A.117I}. An interesting feature that can be observed is the absence of this behavior at depths below $\sim$0.95R$_{\odot}$. 

The flow fields in the QBO timescale drew a very different picture with flow patterns distributed over all latitudes. The amplitudes of the slower-than-average and faster-than average flows increase with increasing depth. Additionally, the flow patterns in the QBO timescale, similar to those in the Schwabe timescale, changes as we go deeper in the solar convection zone. At the depth of 0.99R$_{\odot}$, the flow pattern in the QBO time scale closely resembles that in the Schwabe timescale, forming around $\sim$45$^{\circ}$ and propagate equator- and pole-ward. The deeper layers, on the other hand, exhibit different patterns than those in their Schwabe timescale counterparts.

The amplitude of variations in the Schwabe timescales depends mainly on latitude. Above $\sim$30$^{\circ}$ latitude, the amplitudes increase with increasing latitudes, reaching its maximum above $\sim$55$^{\circ}$ latitudes. Radially, the amplitudes do not exhibit  big variations down to the depth of $\sim$0.78R$_{\odot}$, below which the amplitudes get smaller with increasing depth. Below $\sim$30$^{\circ}$ latitude, the amplitude of variations are almost the same at every latitude with a slight decrease with increasing depth. The amplitude of variations in the QBO, on the contrary, shows primarily radial dependence down to $\sim$0.78R$_{\odot}$, where the amplitude increase with increasing depth, which also was previously observed for solar cycles 23 and 24, separately \citep{2021ApJ...920...49I}. Different from the Schwabe timescale, the maximum amplitude is observed above $\sim$55$^{\circ}$ and below $\sim$0.78R$_{\odot}$, where the amplitudes decrease with decreasing latitude. Considering the slower-than-average flow band coincides with higher magnetic field regions, as a result of the back reaction of the magnetic field on the flow field, one can argue that in the Schwabe time scale, the magnetic field is still mainly confined between $\sim$30$^{\circ}$ N and S latitudes with similar amplitudes. 

To investigate the interaction between the flow and magnetic fields, we assumed that the rotation profiles are symmetric and the magnetic field measured on the surface is non-local. It must be noted that some asymmetries in the flow fields have been shown during declining phase of cycle 23 and rising phase of solar cycle 24 at $\sim$0.99R$_{\odot}$ using local helioseismological inversions \citep{2018ApJ...861..121L}. However, our aim is to have an average picture of the interactions between magnetic and flow fields.

Results from cross-correlation analyses which measure the direction and the strength of the linear relationship at a time lag in the Schwabe timescale show that latitudes between $\sim$30$^{\circ}$ latitude in N and S and below $\sim$0.95R$_{\odot}$ show negative correlations. In this region, the magnetic field leads the flow field. This result is also supported by that from the CCM analyses, showing that the non-linear causal influence is stronger from the magnetic field to the flow field. Above $\sim$0.95R$_{\odot}$ and between $\sim$30$^{\circ}$ latitude in N and S latitudes, however, results from both the cross-correlation and the CCM analyses indicate that there is a positive correlation between the magnetic and flow fields, where the flow field leads the magnetic field and the causal influence of the flow field on the magnetic field is stronger. This region is also where the turbulent $\alpha$-effect is though to be concentrated to generate the observed patterns in active region emergences during a solar cycle \citep{2020LRSP...17....4C}.

Above $\sim$35$^{\circ}$ latitude in N and S, there is a positive correlation between the flow and magnetic fields with a gradual transition from magnetic field leading the flow field to flow field leading the magnetic field with increasing latitude. The only exception for this pattern is the high latitudes in the southern hemisphere, where the magnetic field leads the flow field. The positive correlations are also found to be stronger in the southern hemisphere. An interesting feature is that regions with strong linear positive correlations ($\rho >$ 0.75) coincide with regions where the causal influence of the flow field on the magnetic field is stronger or almost equal. Another interesting feature is the region confined in the northern hemisphere between $\sim$45$^{\circ}$ and $\sim$55$^{\circ}$ and between $\sim$0.75R$_{\odot}$ and $\sim$0.82R$_{\odot}$, where the causal influence of the magnetic field on the flow field is stronger.

In the QBO timescale, on the other hand, the cross-correlation analyses do not exhibit a clear pattern. The negative and positive correlation patches, as well as the lead-lag relationship, are distributed all over latitudes and depths. A similar, chaotic pattern, can also be observed in the causal relationship between the magnetic and flow fields. There are more regions where the causal influence of the flow field on the magnetic field is stronger between $\sim$30$^{\circ}$ latitude in N and S at depths below $\sim$0.95R$_{\odot}$, while above $\sim$30$^{\circ}$ latitude in N and S at all depths, the regions where the causal influence of the magnetic field on the flow field is stronger are more common.

One of the possible explanations for the double cycles is having a secondary dynamo operating in the subsurface region where the strong radial or latitudinal shear exists, while the primary dynamo operating at the bottom of the convection zone where a large-scale radial shear takes place \citep{1998ApJ...509L..49B}. The physical mechanism that produces the high-frequency component is the helicity from the emerged magnetic field of the low-frequency component being imposed on the regions that generates the high-frequency component. Furthermore, the feedback of the low-frequency component on imposed helicity and the amplitude of the high-frequency component show a negative correlation, meaning that the double cycles can be achieved if the two magnetic field sources are weakly interacting. However, it must be noted that this is a simple model that uses only latitudinal shear where cartesian coordinates are employed and the results are therefore qualitative \citep{1998ApJ...509L..49B}.

\citet{2016A&A...589A..56K} used a solar-like semi-global Direct Numerical Simulations (DNS) of convection driven dynamos and showed that the dominant mode exhibiting the most energy around 0.85R$_{\odot}$. This mode is accompanied by a high-and low-frequency components, which operate near the surface around 0.98R$_{\odot}$ and in the bottom of the convection zone near 0.72R$_{\odot}$ \citep{2016A&A...589A..56K}. In addition, results from global 3D nonlinear turbulent MHD simulations, where the feedback of the Lorentz force on the large-scale differential rotation set the cycle-length, showed that there can be two cycles that are non-linearly coupled in the convection zone \citep{2018ApJ...863...35S}. Furthermore, these cycles can display different trends and dependencies, and result from different dynamos. The authors also showed that in their simulations they identified primary cycles from near $\sim$0.72R$_{\odot}$ and secondary cycles originating from the $\sim$0.98R$_{\odot}$ of the convective envelop of their model \citep{2018ApJ...863...35S}. 

Using Sun-as-a-star Doppler velocity observations, \citet{2010ApJ...718L..19F} suggested that the high-frequency component , the QBOs, is an additive contribution to the low-frequency component, the Schwabe cycles, and the amplitude envelope of it is governed by the Schwabe cycle. They also suggested that as the amplitudes of the QBOs seem to be independent of the frequency modes, pointing to a source region deeper than that of the Schwabe cycles  \citep{2010ApJ...718L..19F}, however still within the upper 5\% of the solar radius.

Another explanation for the QBOs comes from the tachocline nonlinear oscillations (TNO) resulting from the interplay between magnetic Rossby waves and differential rotation in the tachocline, leading the toroidal magnetic field to rise buoyantly in the convection zone \citep{2017NatSR...714750D, 2018ApJ...853..144D}. It must be noted that for this interaction to occur, the solar dynamo does not necessarily operate in the tachocline and can be distributed throughout the convection zone \citep{2018ApJ...853..144D}. The physical mechanism that generates the TNOs is the energy exchange between the magnetic Rossby waves and the differential rotation at the solar tachocline. More precisely, magnetic Rossby waves grow and increase their energy by stealing kinetic energy from the differential rotation through Reynolds stress, when the mean shear flow in the tachocline is unstable, this in turn results in the mean flow being tilted agains the shear flow. When Rossby waves reach their maximum kinetic energy, they become weaker as the shear flow cannot provide more energy to the waves, and hence the Rossby waves give back kinetic energy to the mean flow and get tilted with the differential rotation \citep{2017NatSR...714750D,2018ApJ...853..144D}. During the former phase of this oscillatory behaviour, the top layers of the tachocline becomes deformed, creating bulges and depressions, where the tachocline toroidal field starts their buoyant rise in the solar convection zone and form active regions on the solar surface \citep{2017NatSR...714750D,2018ApJ...853..144D}. 

\citet{2021ApJ...920...49I} showed that the QBOs are present in the rotation rate residuals at target depths of 0.99R$_{\odot}$, 0.95R$_{\odot}$, and 0.90R$_{\odot}$, indicating that the source region for the QBOs might extend into the deeper layers of the convection zone, and not confined to the upper 5\% of the solar radius. We also showed here that the amplitude of variations in the rotation rate residuals in the QBO timescale increase with increasing depth down to 0.78R$_{\odot}$, after which the amplitude shows latitudinal dependency, meaning that the amplitude of variations in the QBO timescale increase with increasing latitudes. Our results are more in line with the TNOs as source mechanism for the QBOs, where higher amplitude variations in the rotation rate residuals as we go deeper into the convection zone are expected.

The results from the cross-correlations and CCMs also suggest that the interaction between the rotation rate residuals and magnetic field is not confined to a certain layer in the convective envelope and it is rather distributed with no discernible pattern. For the Schwabe cycle, on the other hand, the source region is distributed across the convection zone and it is confined between $\sim$30$^{\circ}$ latitude in N and S, which are in line with those from the DNS and global 3D simulations. However, we must emphasize that the results from the cross-correlation and CCM analyses must be interpreted with caution because of the assumptions we had to make.

In conclusion, under the assumption that the flow fields are symmetric around the equator and the surface averaged unsigned magnetic field is non-local, the results from our analyses of the interaction between the rotation rate residuals and average unsigned magnetic field strengths in the Schwabe and QBO timescales suggest that the QBOs and Schwabe cycles are originate from a global dynamo mechanism distributed through the convection zone and TNOs are the more likely explanations for the generation of the QBOs.

\acknowledgments

Authors acknowledge computing support from the National Solar Observatory. RH acknowledges support from STFC grant ST/V000500/1. SOHO is a project of international cooperation between ESA and NASA. HMI data courtesy of NASA/SDO and the HMI science team. This work was partially funded as part of the GOES-R Series NASA-NOAA program under the University of Colorado CIRES-NOAA cooperative agreement. The views, opinions, and findings contained in this report are those of the authors and should not be construed as an official National Oceanic and Atmospheric Administration, National Aeronautics and Space Administration, or other U.S. Government position, policy, or decision. 



\bibliography{paper_bibliography}{}
\bibliographystyle{aasjournal}



\end{document}